\documentclass[aps,prd, twocolumn]{revtex4}
\usepackage{graphicx}% Include figure files
\usepackage[hypertex]{hyperref}
\def\beq{\begin{equation}}
\def\eeq{\end{equation}}
\def\bea{\begin{eqnarray}}
\def\eea{\end{eqnarray}}

\def\bwt{\begin{widetext}}
\def\ewt{\end{widetext}}

\def\msun{M_{\odot}}

\def\mpc{{\rm Mpc}}
\def\x{{\bf x}}

\nocite{*}

\usepackage{hyperref}

\begin{document}

\title{Primordial non-gaussianity, statistics of collapsed objects, and the Integrated Sachs-Wolfe effect}
%\author{The Good, et al.}\email{good@good.edu}\affiliation{The Good University, USA}

\author{Niayesh Afshordi}\email{nafshordi@perimeterinstitute.ca}\affiliation{Perimeter Institute
for Theoretical Physics, 31 Caroline St. N., Waterloo, ON, N2L 2Y5,
Canada}
\author{Andrew J. Tolley}\email{atolley@perimeterinstitute.ca}\affiliation{Perimeter Institute
for Theoretical Physics, 31 Caroline St. N., Waterloo, ON, N2L 2Y5,
Canada}
\date{\today}
\preprint{astro-ph/yymmnnn}
\begin{abstract}
Any hint of non-gaussianity in the cosmological initial conditions will provide us with a unique window into the physics of early universe. We show that the impact of a small local primordial non-gaussianity (generated on super-horizon scales) on the statistics of collapsed objects (such as galaxies or clusters) can be approximated by using slightly modified, but {\it gaussian}, initial conditions, which we describe through simple analytic expressions. Given that numerical simulations with gaussian initial conditions are relatively well-studied, this equivalence provides us with a simple tool to predict signatures of primordial non-gaussianity in the statistics of collapsed objects. In particular, we describe the predictions for non-gaussian mass function, and also confirm the recent discovery of a non-local bias on large scales \cite{Dalal:2007cu,Matarrese:2008nc}, as a signature of primordial non-gaussianity. We then study the potential of galaxy surveys to constrain non-gaussianity using their auto-correlation and cross-correlation with the CMB (due to the Integrated Sachs-Wolfe effect), as a function of survey characteristics, and predict that they will eventually yield an accuracy of $\Delta f_{NL} \sim 0.1$ and $3$ respectively, which will be better than or competitive with (but independent of) the best predicted constraints from the CMB. Interestingly, the cross-correlation of CMB and NVSS galaxy survey already shows a hint of a large local primordial non-gaussianity: $f_{NL} = 236 \pm 127$.

\end{abstract}
\maketitle

\section{Introduction}\label{intro}

The standard cosmological model has been tremendously successful in explaining a large range of cosmological observations, ranging from cosmic microwave background (CMB) anisotropies, to the galaxy redshift surveys, supernovae Ia, baryonic acoustic oscillations and the Lyman-$\alpha$ forest in the quasar spectra. However, our clues into the nature of the initial conditions of the Universe, as well as its underlying physics, remain scarce.

The theory of cosmological slow-roll inflation remains the most successful and widely studied model for the cosmological initial conditions. However, despite the large inflow of cosmological observations over the past few years, it remains practically impossible to distinguish a large class of inflationary scenarios. Moreover, given the limited amount of observables accessible to a today's observer, it is not clear whether any possible future observations will be able to falsify the inflationary scenario. One exception to this dilemma would be a possible detection of non-gaussianity  of cosmological initial conditions, which is predicted to be unobservably small in standard slow-roll inflationary scenarios (e.g. \cite{Maldacena:2002vr}). While a large primordial non-gaussianity can still be accommodated in a general inflationary scenario through introducing fine-tuned secondary light fields, higher derivative interactions, or non-adiabatic initial conditions, it could also be interpreted as a smoking gun for alternative early universe models, such as the new ekpyrotic scenario \cite{Buchbinder:2007ad}, which naturally predicts a large primordial non-gaussianity \cite{Buchbinder:2007at}.

The best current observational constraints on the primordial non-gaussianity comes from the statistics of the all-sky CMB maps surveyed by the Wilkinson Microwave Anisotropy Probe (WMAP). Using the 3 year data release of the WMAP CMB maps, Yadav and Wandelt \cite{Yadav:2007yy} claimed a detection of an $f_{NL}$- type local primordial non-gaussianity (see Eq.~(\ref{fnl})), finding $f_{NL} = 87 \pm 30$. This is inconsistent with the expectation from single-field slow-roll inflationary models that predict a much smaller value $|f_{NL}| \lesssim 0.1$, and might be seen as evidence for new ekpyrotic scenario which predicts $f_{NL} \sim 10-100$ \cite{Buchbinder:2007at}. However, the more recent data release of 5 year WMAP maps
shows a smaller value of $f_{NL} = 51 \pm 30$ \cite{Komatsu:2008hk}, which is now consistent with the minimal inflationary expectation at the 2$\sigma$ level. It is thus clear that an accurate determination of the $f_{NL}$ parameter can have important implications for the early universe models. The next generation of all-sky CMB surveys, namely the Planck satellite, is expected to have enough sensitivity to reduce the error on $f_{NL}$ to $\sim 3$ (e.g. \cite{Serra:2008wc}). However, CMB secondary anisotropies (such as Sunyaev-Zel'dovich effect) are expected to contaminate the CMB at this level, and thus prevent a more accurate measurement.

Can we expect complementary (and competitive) constraints from the study
of large scale structure at low redshifts? This question was traditionally addressed by looking at the mass function of galaxy clusters, which was expected to be a sensitive probe of non-gaussian initial conditions (e.g. see \cite{2000ApJ...530...80W,LoVerde:2007ri,Dalal:2007cu} and references therein). Unfortunately, the predicted constraints are not competitive with the CMB, unless $f_{NL}$ has a  strong scale-dependence. Recently, Sefusatti and Komatsu \cite{Sefusatti:2007ih} argued that the analysis of the bispectrum of upcoming large-volume galaxy surveys can yield competitive constraints with the CMB, but their method relies on the accuracy of a local bias model for the distribution of galaxies on small scales. Interestingly, Dalal et al. \cite{Dalal:2007cu} recently predicted the presence of an anomalous large scale power in galaxy distribution as a possible smoking gun for local primordial non-gaussianity (also see \cite{Matarrese:2008nc,McDonald}). This effect would only be seen on very large scales, and thus cannot be confused with any other causal astrophysical effect (although Galactic contamination or large-angle survey systematics could give rise
to a similar spurious power). Shortly before the submission of this paper, Slosar et al. \cite{Slosar:2008hx} published a comprehensive study of this effect in the present large-scale cosmological galaxy surveys, leading to a significant constraint: $f_{NL} = 28 \pm 25$. We will briefly comment on this result in Sec. \ref{conclude}.

In this paper, we present a fully analytic approximation that could recast the statistics of collapsed objects (i.e. galaxies, clusters, etc.) with local primordial non-gaussian initial conditions, in terms of the same statistics with gaussian initial conditions but a modified power spectrum.
In Sec.\ref{linear}, we introduce the approximation that recasts the non-gaussian linear density field in terms of a linear combination of gaussian fields, for collapsing objects of a given mass. Sec.\ref{mass_function} applies this result to the elliptical collapse framework, and compares the resulting mass function with numerical simulations. We then discuss the implications for the large scale clustering of galaxies (due to a non-local bias correction) in Sec.\ref{clustering}, and show that it could yield competitive (or even superior) constraints ($\Delta f_{NL} \sim 0.1$), in comparison with the best upcoming CMB constraints. Finally, Sec.\ref{isw} shows that cross-correlation of CMB with galaxy surveys, due to its sensitivity to the large scale galaxy bias, can similarly put strong constraints on the primordial non-gaussianity which are also competitive with the CMB ($\Delta f_{NL} \sim 3$), and possibly less sensitive to the galaxy survey systematics. Current cross-correlation data between the WMAP CMB maps and the NVSS radio galaxy survey already show a $\sim 2\sigma$ hint for such non-gaussian signal.

Unless mentioned otherwise, we use the WMAP5 best fit $\Lambda$CDM flat cosmology \cite{Komatsu:2008hk} with $\Omega_m = 0.279$, $\Omega_b = 0.0462$, $n_s = 0.96$, $h =0.701$ throughout. For all our numerical calculations, we use the Eisentein \& Hu analytic approximation \cite{Eisenstein:1997ik} for the matter transfer function that includes baryonic features, although we show the simpler BBKS, dark matter only, analytic fit \cite{Bardeen:1985tr} in Eqs.~(\ref{bbks}) \& (\ref{epsilon}), in order to illustrate the asymptotic structure of the transfer function. Finally, all the errors on $f_{NL}$ predicted or quoted in this paper are at 68\% (1$\sigma$) confidence level.

\section{Non-gaussian Linear Perturbations}\label{linear}
The local primordial non-gaussianity has the following form on super-horizon scales:
\beq
\Phi(\x)|_{\rm super}= \Phi_{pG}(\x) - f_{NL} \left[\Phi^2_{pG}(\x)-\langle \Phi_{pG}^2(\x) \rangle\right],\label{fnl}
\eeq
 which was first introduced in \cite{Komatsu:2001rj}, although similar formulations were used in \cite{1991PhST...36..108K,1991MNRAS.248..424M,1992ApJ...390..330S,Gangui:1993tt}. Here $\Phi$ is the longitudinal (or conformal Newtonian) metric perturbation:
\beq
ds^2= (1+2\Phi) dt^2- a^2(t)(1-2\Phi) dx^i dx^i,
 \eeq
and $\Phi_{pG}$ is a statistically homogeneous random gaussian field with a nearly scale-invariant power spectrum. On sub-horizon scales, and during the matter era, the Newtonian potential $\Phi$
is suppressed by the transfer function, $T(k)$, and the metric growth factor $g(t)$ (normalized to unity for $1 \ll z \ll 10^3$) \footnote{$g(t)$ is the growth factor of the metric or Newtonian potential, which is constant and normalized to unity during the matter era, and only starts to decay at the onset of dark energy domination. This should not be confused with $D(z)$, the overdensity growth factor, which grows as $a$ in the matter era, and is often normalized to unity at $z=0$.}:
\beq
\Phi_{\bf k} = T(k) g(t) \Phi_{\bf k}|_{\rm super}.
\eeq
The transfer function can be well approximated by the BBKS fitting formula \cite{Bardeen:1985tr}:
\bea
 &&T(k = q\Omega_m h^2\mpc^{-1}) \simeq \frac{\ln[1+2.34 q]}{2.34 q} \times \\
\nonumber   && \left[1+3.89q+(16.2q)^2+(5.47 q)^3 + (6.71 q)^4\right]^{-1/4},\label{bbks}
\eea

while $g(t)$ only becomes important at the onset of dark energy domination.

%During the matter era, the matter density perturbation, $\delta_m$, is related to $\Phi$ via the $G_0^0$ equation:
Let us {\it define} $\delta_m$ via the Poisson equation

\beq
\delta_{m,{\bf k}} \equiv -\frac{k^2\Phi_{\bf k}}{4\pi Ga^2\rho_m} = -A^{-1}(k;t)\Phi_{\bf k},\label{deltam_def}
\eeq
where
\beq A(k;t)= \frac{3}{2}\Omega_m H_0^2 a(t)^{-1} k^{-2}. \eeq
While $\delta_m$ reduces to the matter overdensity ($=\delta\rho_m/\rho_m$) in the Newtonian limit ($k \gg aH$), the correspondence in the  $k\lesssim aH$ limit will be gauge-dependent, and generally does not hold. However, the determining factor in the dynamics of a spherically collapsing region is its density in units of the local critical density $\Omega_{\rm tot}$. This can be calculated in longitudinal gauge
and within the matter era:
  \beq
  \Omega_{\rm tot} -1 = \frac{10\nabla^2\Phi}{9a^2H^2}.
  \eeq
    This justifies our definition of $\delta_m$ in Eq.~(\ref{deltam_def}), i.e. only the $\nabla^2\Phi$ term (and not the full matter overdensity) is relevant for the statistics of collapsed objects.

Combining the above formulae, we find the following expression for the non-gaussian $\delta_m$ perturbations:
\bea &&\delta_{m,{\bf k}} = \delta _{mG,{\bf k}}+\\ \nonumber && \frac{f_{NL}}{g(t)}\int \frac{d^3k'}{(2\pi)^3} \frac{A(|{\bf k-k'}|;t)A(k';t) T(k)}{T(|{\bf k-k'}|)T(k') A(k;t)} \delta_{mG,{\bf k -k'}}\delta_{mG,{\bf k'}}, \label{deltang}\eea where
\beq
\delta_{mG,\bf k} = -A^{-1}(k;t) T(k) g(t) \Phi_{pG,{\bf k}},
\eeq
is the gaussian density perturbation (if $f_{NL} =0$).
In order to find a local form for $\delta_m$ in real space,
let us consider the approximation:
\bea
\frac{A(|{\bf k-k'}|;t)A(k';t) T(k)}{T(|{\bf k-k'}|)T(k') A(k;t)} \simeq \nonumber\\ \frac{A(k';t)}{T(k')}+\frac{A(|{\bf k-k'}|;t)}{T(|{\bf k-k'}|)}-\frac{A(k;t)}{T(k)}. \label{approx}
\eea
Assuming $k,k' \gg k_{eq}$ (associated with collapsed objects today), notice that for large $k$'s
$A(k;t)/T(k)$ only logarithmically depends on $k$. Therefore, since Eq.~(\ref{approx}) holds for constant $A/T$, it must be a good approximation in the large $k$ limit.
The only exception to this limit is when $k' \ll k_{eq}$ ($|{\bf k - k'}| \ll k_{eq}$), for which the first (second) term on the right hand side dominates ($A/T \propto k^{-2}$ for small $k$'s). However, since the two other terms are still slowly varying, Eq.~(\ref{approx}) remains a good approximation. In fact, one can numerically confirm that Eq.~(\ref{approx}) remains accurate at the percent level for $k' \lesssim k$.

%Let's look at the different limits of the function $T(k)/A(k;t)$.% In this regime, A(k;t) takes the simple form:
%For the transfer function, if $k \ll k_{eq} \sim 0.2 \times\Omega_m h^2 \mpc^{-1}$, $T(k) \simeq 1$, while for $k \gg k_{eq}$ we have:
%\beq
%T(k) \simeq 1.00 \times 10^{-3} \frac{\ln\left[18.28~ k(\mpc^{-1})\right]}{k^2(\mpc^{-1})},
%\eeq
%where we used $\Omega_m h^2 \simeq 0.128$. Therefore, in the limit $k \gg k_{eq}$:
%\beq \frac{T(k)}{A(k;t)} \simeq 4.7 \times 10^{4} a(t) \ln\left[18.28~ k(\mpc^{-1})\right]. \eeq
%Therefore, in the large $k,k'$ limit, the pre-factor in the integrand of Eq.~(\ref{deltang}):
% \beq
% \frac{10}{9g(t)}\frac{A(|{\bf k-k'}|;t)A(k';t) T(k)}{T(|{\bf k-k'}|)T(k') A(k;t)} \simeq \frac{2.4 \times 10^{-5}\ln\left[18.28~ k(\mpc^{-1})\right]}{a(t)g(t)\ln\left[18.28~ k'(\mpc^{-1})\right]\ln\left[18.28~ |{\bf k-k'}|(\mpc^{-1})\right]} \simeq \epsilon(M) (ag)^{-1},
% \label{prefactor}\eeq

Plugging Eq.~(\ref{approx}) into Eq.~(\ref{deltang}), we arrive at the following local approximation for the linear non-gaussian overdensity in the real space:
 \bea
  &&\delta_{m}(\x;t,M) \simeq \delta_{mG}(\x;t,M) - f_{NL} \times \nonumber\\ &&\left\{2\Phi_{pG}(\x;M) \delta_{mG}(\x;t,M)+ \frac{\epsilon(M)}{ag}\left[\delta_{mG}(\x;t,M)\right]^2\right\},\nonumber\\\label{deltax}
  \eea
 where, for collapsed haloes of mass $M$, we assumed $k/\pi \sim R^{-1}(M) \equiv [4\pi\bar{\rho}_m/(3M)]^{1/3}$, and thus
 \bea
 &&\epsilon(M) \equiv \frac{a(t) A(k;t)}{T(k)} \simeq \frac{1.9\times 10^{-5} }{(\Omega_m h^2/0.137)\ln(1+2.34q)} \times \nonumber\\ &&\left[1+0.081q^{-1}+0.129q^{-2}+0.00192q^{-3}+0.00049q^{-4}\right]^{1/4}, \nonumber\\ &&{\rm with } \quad q=\left(M\over 1.9\times 10^{15} \msun\right)^{-1/3}\left(\Omega_m h^2\over 0.137\right)^{-2/3}.\label{epsilon}
 \eea

A second approximation that we made in writing Eq.~(\ref{deltax}) was to ignore the contribution of gaussian modes with wavelength smaller than the mass filter ($k > \pi/R(M)$) to the quadratic term. This is justified as the power {\it per mode} drops as $\sim k^{-3}$ on small scales, and thus the constructive interference of short wavelength modes in the quadratic term cannot significantly change the density averaged over the mass filter. In other words, the contribution of the Fourier modes to the total power in the non-gaussian term (averaged over the mass filter) is sharply cut off beyond $k \sim \pi/R(M)$, which justifies using the smoothed gaussian fields on the right hand side of Eq.~(\ref{deltax}).

%Thus, in this limit ($q\gtrsim 1$), Eq.~(\ref{deltang}) becomes a simple convolution in Fourier, and so a local equation in real space.
%
%However, even if $k \gg k_{eq}$,  $k'$ or $|{\bf k-k'}|$ could be smaller than $k_{eq}$. For $k' \lesssim k_{eq} \ll k$, one can safely assume $|{\bf k-k'}| \simeq k$, implying that the pre-factor in Eq.~(\ref{prefactor}) grows as $k'^{-2}$. In this limit, we get:
% \beq
% \frac{10}{9g(t)}\frac{A(|{\bf k-k'}|;t)A(k';t) T(k)}{T(|{\bf k-k'}|)T(k') A(k;t)} \delta^G _{m,{\bf k -k'}}\delta^G _{m,{\bf k'}} \simeq -\delta^G_{\bf k-k'} \Phi_{pG,{\bf k'}}, \eeq
%A similar expression can be found in the $|{\bf k-k'}| \lesssim k_{eq} \ll k$ limit, which becomes identical to the above expression under the change of variable ${\bf k'} \rightarrow {\bf k - k'}$. Finally, combining these three regimes, we find:
%\beq
%  \delta_{m}(\x) \simeq \delta^G _{m}(\x) + f_{NL}\left\{-\frac{20}{9g}\Phi_G(\x) \delta^G _{m}(\x)+ \frac{\epsilon(M)}{ag}\left[\delta^G _{m}(\x)\right]^2\right\},\label{deltax}
%  \eeq
%  where we used $\Phi_{G,{\bf k'}} \equiv -A(k';t)\delta^G_{m,{\bf k'}}$. Notice that, since the $\Phi_G$ term is only for $k' \lesssim k_{eq}$ regime, $T(k') \simeq 1$, and so we could replace $\Phi_{pG}$ by $10\Phi_G/g(t)$.
%
%  With the exception of the very early structures, the last term in Eq.~(\ref{deltax}) is $ \lesssim 0.1\%$ of the first (gaussian) term, and so is negligible for structure formation. Therefore, we can safely ignore it, to find:
%  \beq \delta_{m}(\x) \simeq \left[1- \frac{20f_{NL}}{9g}\Phi_G(\x)\right]\delta^G _{m}(\x).\eeq
\section{Statistics of Collapsed Objects}\label{mass_function}

  In order to estimate the impact of non-gaussianity on the statistics of collapsed objects, we will use the ellipsoidal collapse framework \cite{Sheth:1999su}, where the collapse criterion is that $\delta_m(\x;t,M)$ (smoothed with a top-hat filter of mass $M$) exceeds the mass-dependent collapse threshold $\delta_{ec}(M)$
  \beq
\delta_{ec}(M) \simeq q^{1/2}\delta_{sc} \left[1+\beta (q\nu)^{-\gamma}\right], \nu = \left[\delta_{sc}\over\sigma(M)\right]^2,\label{ecol}
  \eeq
  where $(q,\beta,\gamma)=(0.707,0.47,0.615)$ are fitted to simulations \cite{Sheth:2001dp}, and $\delta_{sc} \simeq 1.68$ is the spherical collapse threshold.

  For gaussian initial conditions, numerical simulations indicate that, to a good approximation, the mass function (or number density) of collapsed objects of mass $M$ is simply a function of $\sigma(M)/\delta_{ec}(M)$, where $\sigma^2(M) = \langle \delta_{mG}^2(\x;t,M) \rangle$.

  \begin{figure}
\includegraphics[width=\columnwidth]{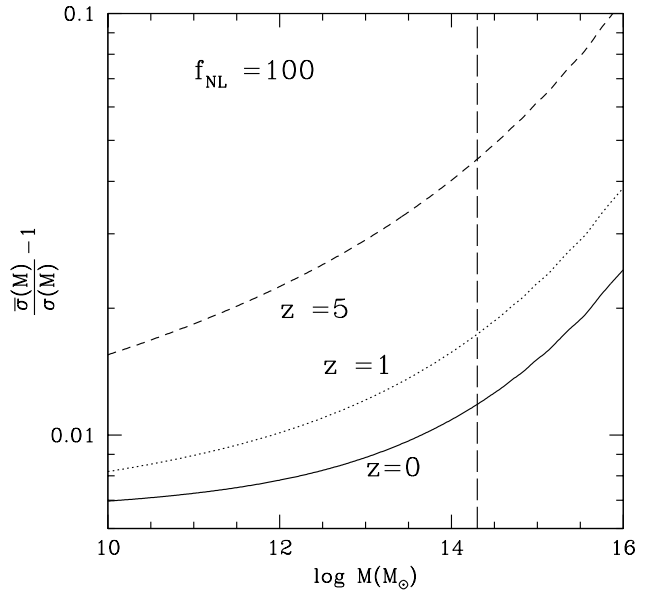}
\caption{The non-gaussian correction to $\bar{\sigma}(M)$, the {\it effective} amplitude of density fluctuations for collapsed objects of mass $M$. The solid, dotted, and dashed lines are for $z=0,1,5$ respectively, while we assumed $f_{NL}=100$. The vertical line shows $M_8$, the mass within comoving $8 h^{-1}{\rm Mpc}$. } \label{sigma_bar}
\end{figure}

  On the other hand, Eq.~(\ref{deltax}) shows that, for the non-gaussian initial conditions, the local overdensity depends on the local values of both $\delta_{mG}$ and $\Phi_{pG}$. To linear order in $f_{NL}$, the elliptical collapse condition reduces to:
  \beq
  \delta_m > \delta_{ec} \Rightarrow \delta_{mG} -2f_{NL}\delta_{ec}\Phi_{pG} \gtrsim \delta_{ec} + \frac{\epsilon f_{NL}}{ag} \delta_{ec}^2.
  \eeq
  We can now define $\tilde{\delta}_{mG}$ and $\tilde{\delta}_{ec}$ as:
  \beq
  \tilde{\delta}_{mG} \equiv  \delta_{mG} -2f_{NL}\delta_{ec}\Phi_{pG}, \quad \tilde{\delta}_{ec} \equiv \delta_{ec} + \frac{\epsilon f_{NL}}{ag} \delta_{ec}^2.\label{delta_tilde}
  \eeq

  We thus find that, as $\tilde{\delta}_{mG}$ is a gaussian random field, we could still use the old gaussian framework, provided we replace
  $\delta_{ec}$ by a slightly modified collapse threshold $\tilde{\delta}_{ec}$. Moreover, as the mass function only depends on the ratio $\tilde{\sigma}(M)/\tilde{\delta}_{ec}$, we have:
  \beq
  \frac{\tilde{\sigma}(M)}{\tilde{\delta}_{ec}} =  \frac{\langle\tilde{\delta}_{mG}\rangle^{1/2}}{\tilde{\delta}_{ec}}= \frac{\bar{\sigma}(M)}{\delta_{ec}},
  \eeq
  where
   \bea
  &&\bar{\sigma}(M) \equiv \tilde{\sigma}(M) \frac{\delta_{ec}}{\tilde{\delta}_{ec}} \nonumber\\ &\simeq&  \left[\sigma^2(M)-4f_{NL}\delta_{ec}\langle\Phi_{pG}\delta_{mG}\rangle \right]^{1/2} \left(1-\frac{\epsilon f_{NL}}{ag} \delta_{ec}\right)  \nonumber\\ &\simeq& \sigma(M) \left[1-\frac{f_{NL}\delta_{ec}}{ag}\left(\frac{2ag\langle\Phi_{pG}\delta_{mG}\rangle}{\sigma^2(M)}+\epsilon \right)\right] .\label{sigma_bar_eq}
  \eea
   We also note that:
   \bea
   \sigma^2(M) &=& \int \frac{d^3k}{(2\pi)^3} P(k) W^2(kR),\\
    ag\langle\Phi_{pG}\delta_{mG}\rangle &=& -a\int \frac{d^3k}{(2\pi)^3} \frac{A(k)}{T(k)}P(k) W^2(kR),
    \eea
    where, $W(kR)$ is the Fourier transform of the spherical top-hat filter of radius $R$:
    \beq
     W(x)= \frac{3(\sin x-x\cos x)}{x^3},
    \eeq
    and
    $
    4\pi R^3\bar{\rho}_m/3 = M,
    $
    while $\epsilon$ was defined in Eq.~(\ref{epsilon}).
    %In particular, notice that for small masses or $k_{eq}R \ll 1$, we asymptotically have:
%    \beq
%     \frac{2ag\langle\Phi_{pG}\delta_{mG}\rangle}{\sigma^2(M)}+\epsilon  \rightarrow -\epsilon,
%    \eeq

    Fig.~(\ref{sigma_bar}) shows the non-gaussian correction to $\bar{\sigma}(M)$  for $f_{NL} = 100$ as described in Eq.~(\ref{sigma_bar_eq}), for three different redshifts. We first notice the increasing trend with mass, i.e. the effective non-gaussian amplification of the power spectrum is more significant for larger haloes/clusters. Secondly, we notice the increasing trend with redshift, which simply results from the inverse scaling with $ag$ in Eq.~(\ref{sigma_bar_eq}). The change in the shape of the correction at different redshifts, however, is a result of the mass dependence of the elliptical collapse threshold (Eq.~\ref{ecol}), which kicks in below the (redshift-dependent) non-linear mass scale.

    The vertical line shows $M_8$, the mass within comoving $8 h^{-1}{\rm Mpc}$, which is often used to normalize the amplitude of linear density fluctuations: $\sigma_8 = \sigma(M_8)$. For our assumed cosmology, we find approximately
    \bea
    &&\frac{\Delta\sigma_8}{\sigma_8} \simeq 1.21 \times 10^{-2} D(z)^{-0.8}\left(f_{NL}\over 100\right) \left(\Omega_m h^2 \over 0.137\right)^{-1},\label{dsigma8}\\
    &&\Delta n_8 \simeq 7.2 \times 10^{-3} D(z)^{-1.2} \left(f_{NL}\over 100\right) \left(\Omega_m h^2 \over 0.137\right)^{-1},\label{dn8}
    \eea
    where $\Delta\sigma_8$ and $\Delta n_8$ are corrections to the effective normalization and tilt of the matter power spectrum at $8 h^{-1}{\rm Mpc}$, {\it if} it is deduced from the statistics of collapsed objects at redshift $z$, using gaussian predictions. Here, $D(z) \propto ag$ is the linear density growth factor, which is normalized to unity ar $z=0$, and we used the definition \beq\frac{n_{\rm eff}+3}{6} = -\frac{d\ln\bar{\sigma}(M)}{d\ln M}.\eeq

    One implication of this result is that, assuming constant $f_{NL} <100$, one needs to measure $\sigma_8$ to better than $1\%$, in order to be able to distinguish gaussian and non-gaussian models from cluster number counts. However, assuming that this is the case, the degeneracy with $\sigma_8$ can be broken through the redshift evolution of the mass function (or the effective power spectrum).

 \begin{figure}
\includegraphics[width=\columnwidth]{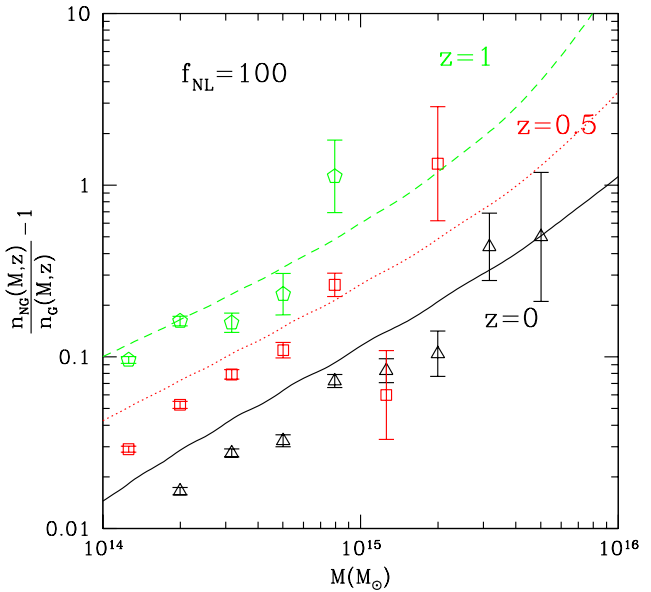}
\caption{The relative change in the cluster mass function for $f_{NL} = 100$. The solid (black), dotted (red), and dashed (green) lines correspond to our analytic prediction based on the $\sigma(M)$ correction (Eq.~(\ref{dsigma8}-\ref{dn8})) for $z=0,0.5$ and $1$ respectively. The triangles, squares, and pentagons are simulations of Dalal et al. \cite{Dalal:2007cu} in the same order.
The error bars show simple Poisson errors in simulated cluster numbers.} \label{dn}
\end{figure}
%Now, using Eq.~(\ref{delta_ls}), the elliptical collapse threshold for $\delta_{mG,S}$ becomes:
%  \beq
%    \delta_{ec,G}(M) \simeq  \delta_{ec}(M) - \delta_{mG,L} + f_{NL}\delta_{ec}(M)\left\{\Phi_{pG,L} -\frac{\epsilon(M)}{ag}\left[\delta_{ec}(M)- \delta_{mG,L}\right]  \right\}.
%    \eeq
%  Notice that, in using Eq.~(\ref{delta_ls}), we have replaced the smoothed $\delta_{mG,S}^2$ (on the mass scale $M$), by the square of the smoothed $\delta_{mG,S}$, which is certainly an underestimate. However, since linear density amplitude only logarithmically grows on small scales, we expect this approximation to be reasonably accurate.
%
%  Next, we note that simulations with {\it gaussian} initial conditions show the mean halo density is well described by a universal profile:
%  \beq
%  n_{\rm halo}(M)dM = \frac{\bar{\rho}_m}{M} f(\nu) d\nu,
%  \eeq
%  where $\nu$ was defined in Eq.~(\ref{ecol}).

 Fig.~(\ref{dn}) compares the predictions of our analytic framework for the cluster mass function with the simulated mass functions with non-gaussian initial conditions. In order to find the mass function, we plug the effective $\sigma(M)$ (Eqs.~(\ref{sigma_bar_eq}) or (\ref{dsigma8}-\ref{dn8})) into the Warren et al. analytic formula \cite{Warren:2005ey} for the halo mass function, which was fitted to $\Lambda$CDM N-body simulations with gaussian initial conditions. This result is compared  with the mass functions directly obtained in Dalal et al. \cite{Dalal:2007cu} using non-gaussian initial conditions. The analytic prediction uses the same cosmology assumed in the non-gaussian simulations

 It appears that our analytic framework can reproduce the numerical results in the large mass regime (when the corrections are $>10\%$), but overpredicts the mass function by up to $\sim 50\%$ at lower masses, although the discrepancy seems less pronounced at higher redshifts. One possible reason for this discrepancy is that the elliptical collapse threshold of Eq.~(\ref{ecol}) \cite{Sheth:1999su} is based on an analytic fit to random gaussian fields. Using non-gaussian initial conditions will most likely introduce corrections to the elliptical collapse threshold, which could become significant at small masses (although for large masses one still expects the spherical collapse threshold $\delta_{sc} \simeq 1.68$).

 Another possibility is the presence of an unknown systematic error (or numerical artifact) in the non-gaussian simulations. For example, it is clear from the simulation data points in Fig.~(\ref{dn}) that the Poisson errors are only an underestimate for the true numerical error. Moreover, it appears that Dalal et al. simulations \cite{Dalal:2007cu} predict a lower non-gaussian correction to the mass function, in comparison with independent simulations of Grossi et al. \cite{Grossi:2007ry}.

 Alternatively, one may argue that such a discrepancy at low masses,
 even if real, may be of little consequence as non-gaussian corrections are only large (and thus observable) for massive clusters. Nevertheless, more thorough numerical studies (including a suite of non-gaussian simulations to account for sample variance), as well as an extension of the elliptical collapse framework to non-gaussian initial conditions should elucidate this issue.

\section{Clustering on large scales}\label{clustering}

The clustering of collapsed objects on large scales is often described in terms of the bias parameter, which quantifies the ratio of the overdensity of collapsed objects to the total matter overdensity:

\beq
b(k,M) \equiv \frac{\langle\delta_{m,{\bf k}} \delta_{M,-{\bf k}}\rangle}{\langle\delta_{m,{\bf k}} \delta_{m,-{\bf k}}\rangle},
\eeq

where
\beq
\delta_{M,{\bf k}} = \langle n_{\rm halo}(M)\rangle ^{-1}\int d^3{\bf x} e^{-i{\bf k\cdot x}} \delta n_{\rm halo}({\bf x},M),
\eeq
is the Fourier transform of the overdensity of haloes of mass $M$.

In the last section, we showed that the statistics of collapsed objects of mass $M$, resulting from (slightly) non-gaussian initial conditions, can be described in terms of the statistics of a new gaussian field $\tilde{\delta}_{mG}$, with a modified elliptical collapse threshold $\tilde{\delta}_{ec}$ (Eq.~\ref{delta_tilde}). In particular, the fraction of the mass in collapsed objects larger than mass $M$ is given by:
\beq
f(>M) = F(x); \quad x \equiv {\tilde{\delta}_{ec}(M) \over \tilde{\sigma}(M)} = {\delta_{ec}(M) \over \bar{\sigma}(M)}, \label{fgm}
\eeq
and $F(x)$ is a universal function that can be fitted from simulations, and roughly has the Press-Schecter form:
\beq
F(x) \simeq F_{PS}(x) = \sqrt{\frac{2}{\pi}}\int^{\infty}_{x} e^{-y^2/2} dy.\label{PS}
\eeq
Similarly, the fraction of mass in the halo mass interval $(M,M+dM)$ is given by:
\beq
f(M,M+dM) = -F'(x)\left(\frac{\tilde{\delta}_{ec}(M+dM)}{\tilde{\sigma}(M+dM)} -  \frac{\tilde{\delta}_{ec}(M)}{\tilde{\sigma}(M)}\right)\label{fm}
\eeq

Now, adding a positive background of $\tilde{\delta}_0$ ($=\delta_{mG,0}-2f_{NL}\delta_{ec}\Phi_{pG,0}$) to $\tilde{\delta}_{mG}$ in a large patch of the Universe effectively decreases the collapse threshold to $\tilde{\delta}_{ec} - \tilde{\delta}_0$ for small scale fluctuations in that region. Therefore, the collapsed mass fraction in that region is slightly modified:
\beq
\delta f(M,M+dM) = \tilde{\delta}_0\left\{{F''(x)\over \tilde{\sigma}(M)}d\left[\tilde{\delta}_{ec}(M)\over\tilde{\sigma}(M)\right]+F'(x) d\tilde{\sigma}^{-1}(M)\right\}. \label{dfm}
\eeq
Combining Eqs. (\ref{fm}) and (\ref{dfm}), and after simple manipulations, we find:
\beq
\frac{\delta f}{f}(M,M+dM) = \frac{\tilde{\delta}_0}{\tilde{\delta}_{ec}(M)}\left\{-\frac{d\ln F'}{d\ln x} + \left[ \frac{d\ln \tilde{\delta}_{ec}(M)}{d\ln \tilde{\sigma}(M)}-1\right]^{-1}\right\},
\eeq
which is also known as $Lagranigan$ overdensity. Since the mean halo number density is given by $\bar{\rho}_m f(M,M+dM)$, the (Eulerian) halo overdensity with the $\tilde{\delta}_0$ background takes the form:
\beq
\delta_{\rm halo} = \delta_{mG0} + \frac{\tilde{\delta}_0}{\tilde{\delta}_{ec}(M)}\left\{-\frac{d\ln F'}{d\ln x} + \left[ \frac{d\ln \tilde{\delta}_{ec}(M)}{d\ln \tilde{\sigma}(M)}-1\right]^{-1}\right\}.
\eeq

Recalling the definition of $\tilde{\delta}$, this yields the following expression for the (Eulerian) bias:
\bwt
\beq
b(k,M)=1+\tilde{\delta}_{ec}(M)^{-1}\left[1+{2f_{NL}\delta_{ec}(M)A(k,t)\over g(t)T(k)}\right]\left\{-\frac{d\ln F'}{d\ln x} + \left[ \frac{d\ln \tilde{\delta}_{ec}(M)}{d\ln \tilde{\sigma}(M)}-1\right]^{-1}\right\},\label{bias_gen}
\eeq
or
\beq
b(k,M) = b_{\infty}(M) + \frac{2(b_{\infty}-1)f_{NL}\delta_{ec}(M)A(k,t)}{g(t)T(k)} = b_{\infty}(M) + \frac{3(b_{\infty}-1)f_{NL}\delta_{ec}(M)\Omega_mH^2_0}{k^2a(t) g(t)T(k)},\label{bias_ng}
\eeq
\ewt

 \begin{figure}
\includegraphics[width=\columnwidth]{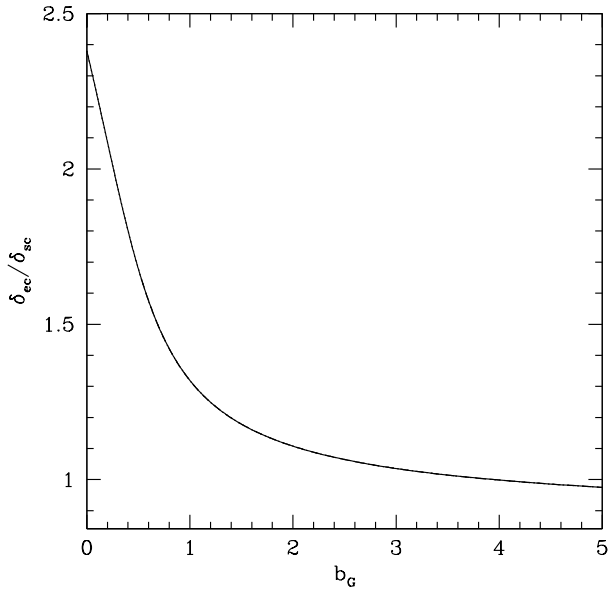}
\caption{The ratio of the elliptical to spherical collapse thresholds, as a function of the scale-independent gaussian linear bias. The scale-dependent non-gaussian bias is enhanced by this factor (Eq.~\ref{bias_ng}).} \label{deltaec}
\end{figure}

Here $b_{\infty}(M) = b(M;f_{NL}=0) + {\cal O}\left(f_{NL} \epsilon\over ag\right) $ is the $k \rightarrow \infty$ limit of the {\it linear} bias. For reasonable values of (constant) $f_{NL} < 100$, similar to $\bar{\sigma}(M)$, this would only lead to percent level corrections to the gaussian bias, which we can safely ignore and assume $b_{\infty}(M) \simeq b(M;f_{NL}=0)$.  With this assumption, the ratio of the elliptical to spherical collapse threshold, $\delta_{ec}(M)/\delta_{sc}$, (again ignoring $f_{NL}$) can also be described as a function of $b_{\infty}$ by combining Eqs. (\ref{ecol}), (\ref{PS}), and (\ref{bias_gen}). This is plotted in Fig.~(\ref{deltaec}), and can be used to estimate the non-gaussian scale-dependent bias (second term in Eq.~\ref{bias_ng}) as a function of the small-scale (or gaussian) linear bias.

\begin{figure}
\includegraphics[width=\columnwidth]{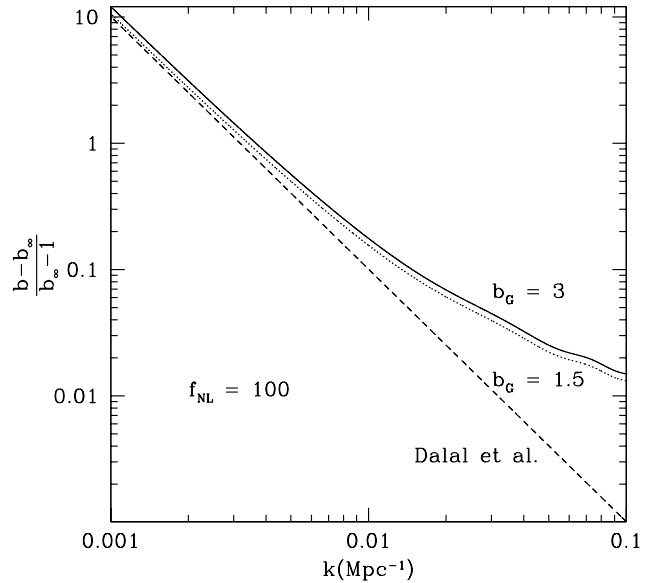}
\caption{The scale-dependent part of bias $b-b_{\infty}$ divided by $b_{\infty}-1$, for $f_{NL}=100$ and $z=0$. The scale-independent part of bias, $b_{\infty} = 1.5$ and $3$ for solid and dotted curves. The dashed line is Dalal et al.'s derivation \cite{Dalal:2007cu}.  \label{ngauss_bias}}
\end{figure}
Fig.~(\ref{ngauss_bias}) shows the scale-dependent part of linear bias (the second term in Eq.~\ref{bias_ng}) as a function of $k$ for $b_{\infty}=1.5,3$ and $f_{NL}=100$, which is compared to the Dalal et al. derivation \cite{Dalal:2007cu}. For small $k$'s
($k<k_{eq}$), both biases scale as $k^{-2}$, although our prediction is slightly higher by the elliptical to spherical collapse ratio (Fig.~(\ref{deltaec})). However, for $k \gtrsim k_{eq}$ the non-gaussian scale-dependent bias factor is further enhanced by the inverse of the transfer function, which was missing in \cite{Dalal:2007cu} (and was also recently pointed out by \cite{Matarrese:2008nc}).

%  and thus the ellipsoidal collapse threshold (in terms of the gaussian overdensity) is lowered to:
%  \beq \delta_{mG}(\x) \gtrsim \tilde{\delta}_c(M) = \delta_c(M)-\bar{\delta}_m + \frac{20}{9}f_{NL}g^{-1}\delta_c(M)\Phi_G(\x). \eeq
%  Since $\Phi_G(\x)$ only varies on large scales, we can use the same mass function with a shifted collapse threshold, $\tilde{\delta}_c$:
%  \beq \ln \frac{dn}{d\ln[\sigma^{-1}(M)]} \rightarrow \ln \frac{dn}{d\ln[\sigma^{-1}(M)]} + \bar{\delta}_m + \frac{(\nu^2-1)}{\delta_c} \left[\bar{\delta}_m -\frac{20}{9}f_{NL}g^{-1}\delta_c(M)\Phi_G(\x)\right],\eeq or equivalently \beq \delta_{\rm halo}(\x) = b(M)\bar{\delta}_m(\x) - \frac{20}{9g}f_{NL}[b(M)-1]\delta_c(M) \Phi_G(\x), \eeq
%  where
%  $b(M)= 1+\frac{\nu^2-1}{\delta_c(M)}$ is the Mo \& White halo bias.
%

\begin{figure}
\includegraphics[angle=-90,width=\columnwidth]{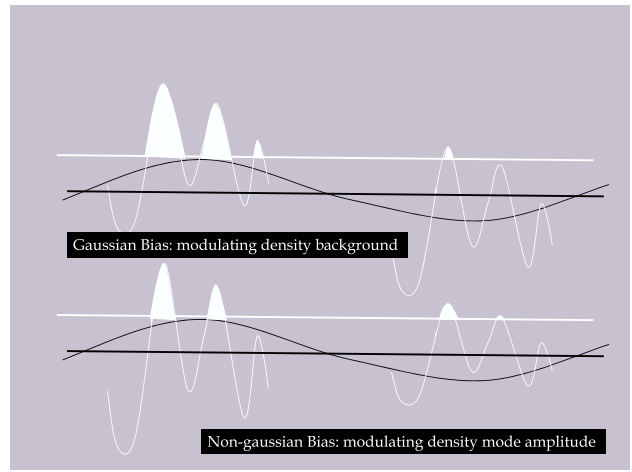}
\caption{This figure illustrates the contrast between the gaussian and non-gaussian halo/galaxy bias. The two plots show cartoon versions of linear density vs. spatial position. The white shaded areas indicate collapse regions. For gaussian initial conditions, different Fourier modes are uncorrelated, and so long wavelength modes (thin black curve) only change the background local mean value of small scale modes (thin white curves), which in turn changes the number of density peaks that cross the collapse threshold (thick white line) and form haloes. However, for non-gaussian initial condition, the long wavelength modes can also modulate the amplitude of small scale modes, which causes an additional modulation of collapsed halo density.  \label{ngauss_illus}}
\end{figure}

The unique aspect of this non-gaussian correction to the galaxy bias is that, rather than going to a constant on large scales, it blows up as $k \rightarrow 0$. This is due to the fact that, on large scales, $\tilde{\delta}_{mG}$ is dominated by the $\Phi_{pG}$ term (see Eq.~\ref{delta_tilde}), and thus, the galaxy distribution follows the primordial curvature perturbations rather than the matter density.
An easy way to understand this phenomenon is illustrated in Fig.~(\ref{ngauss_illus}), which shows how modulation of the amplitude of small scale density perturbations by large scale modes (caused by primordial non-gaussianity) can lead to an additional non-gaussian bias.

\begin{figure*}
\includegraphics[width=0.65\columnwidth]{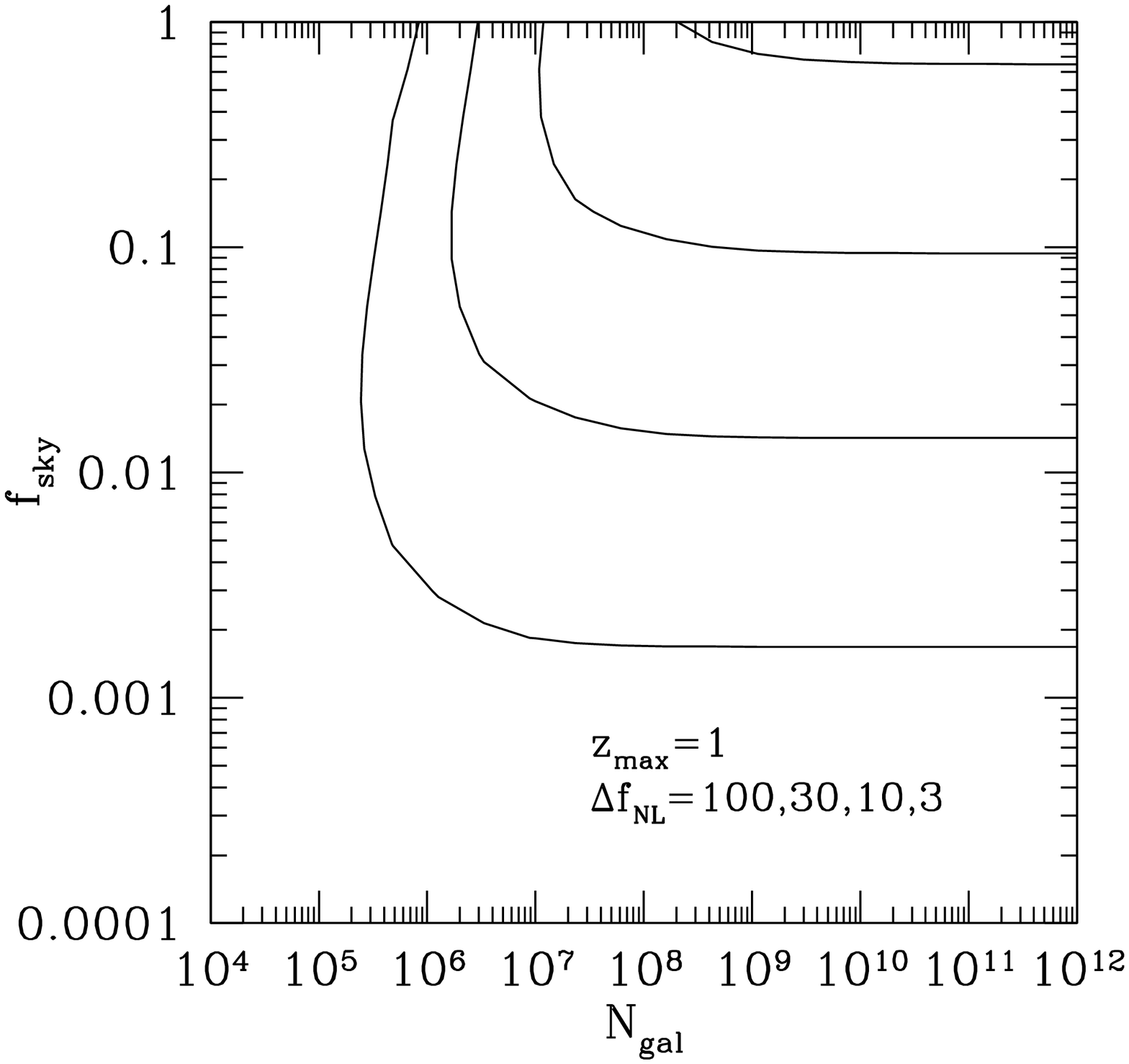}
\includegraphics[width=0.65\columnwidth]{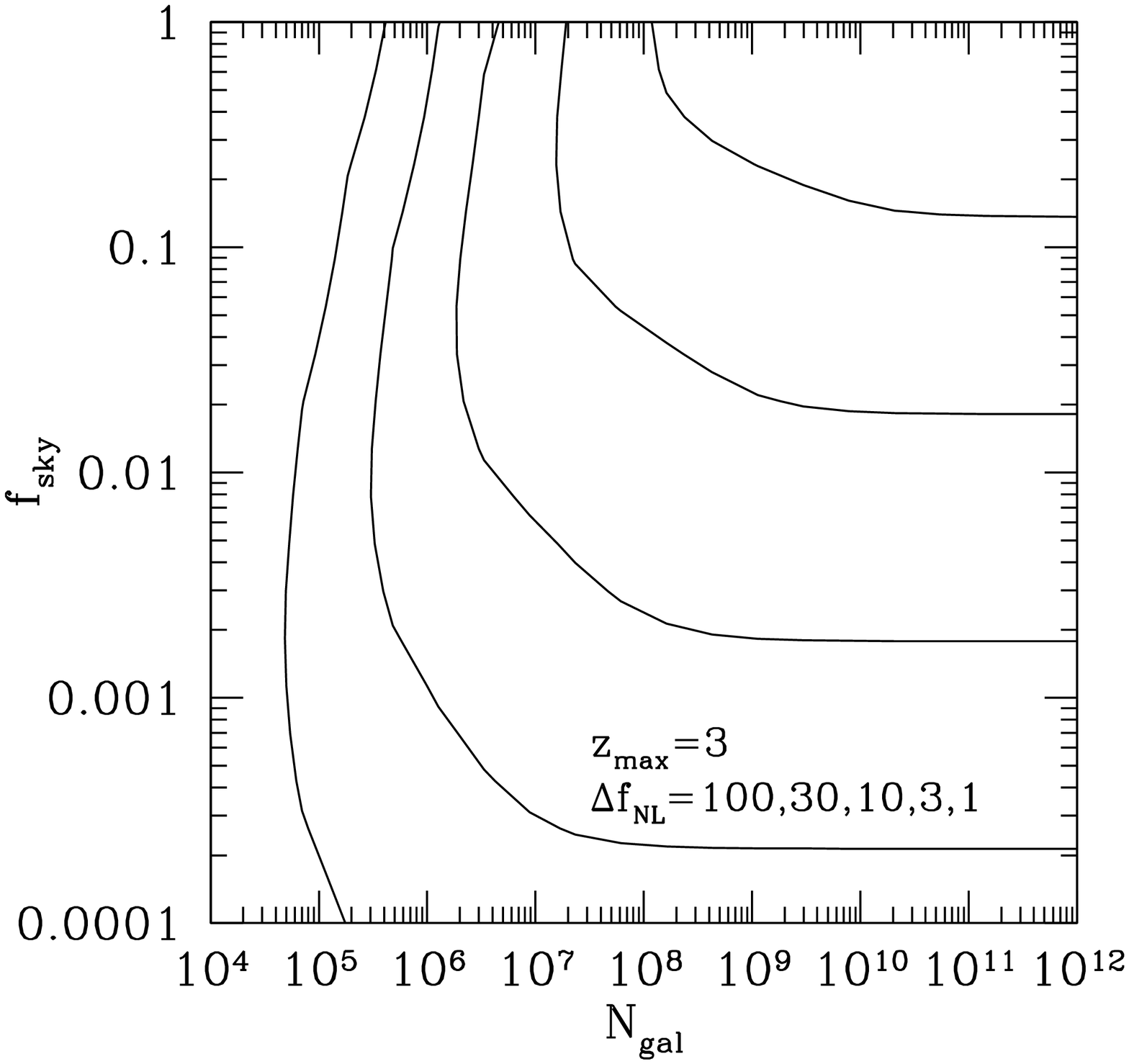}
\includegraphics[width=0.65\columnwidth]{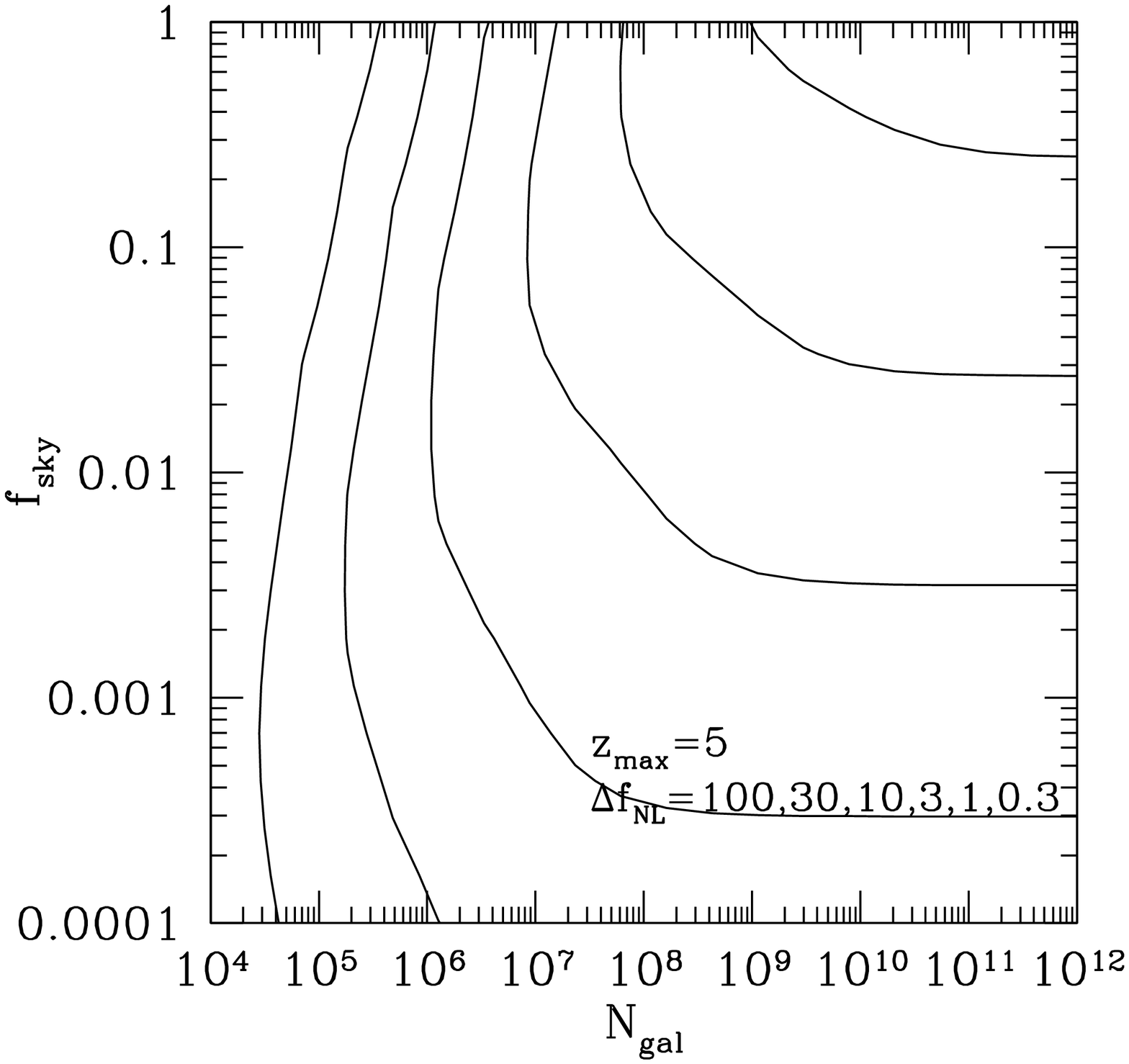}
\caption{The expected precision for the measurement of local primordial non-gaussianity, using the auto-power spectrum of a galaxy survey (in lieu of any survey systematics). The contours show the expected precision as a function of the survey sky coverage, $f_{sky}$, and the number of galaxies in the survey, $N_{gal}$ (assuming uniform comoving galaxy number density and $b_{gal}=2$), for $z_{max}=1,3,5$. \label{dfnl}}
\end{figure*}

The peculiar scale-dependence of the non-gaussian bias correction differentiates it from any causal astrophysical effect, where bias should go to a constant on large scales (beyond the sound horizon $\sim$ few $\mpc$'s). Therefore, this effect can provide us with a smoking gun for primordial non-gaussianity, through observations of large scale clustering in galaxy surveys \cite{Dalal:2007cu, Matarrese:2008nc}. In fact, Dalal et al. \cite{Dalal:2007cu} claim that next generation of large scale galaxy surveys can yield $\Delta f_{NL} \lesssim 10$. In Fig.~(\ref{dfnl}) we illustrate this potential for a galaxy redshift survey with a given number of galaxies $N_{gal}$, fraction of sky coverage $f_{sky}$, and maximum redshift $z_{max}$. Here, for simplicity, we assume a uniform comoving galaxy number density, and $b_{gal} =2$ for the Gaussian bias. In order to compute the sensitivity of the observation of galaxy power spectrum to an $f_{NL}$-type primordial non-gaussianity, we consider the gaussian $\chi^2$:
\beq \chi^2 = \int dV(z) \int^{k_{max}}_{k_{min}} \frac{d^3k}{(2\pi)^3} \frac{|\delta_{gal}({\bf k})|^2}{P_{gal}(k,z)}+\ln[P_{gal}(k,z)], \label{chi2_survey}\eeq
where $P_{gal}(k,z)$ is the power spectrum of galaxy distribution, which, in the linear regime, is related to the matter power spectrum, $P(k,z)$, through bias and the comoving galaxy number density: \beq P_{gal}(k,z) = b_{gal}^2(k,z) P(k,z) + n^{-1}_{gal}. \eeq
$b_{gal}(k)$ is given by Eq.~(\ref{bias_ng}), which is parameterized by $f_{NL}$ and $b_{\infty}$.  Assuming that these parameters remain constant across the survey, they can be constrained observationally through minimizing the $\chi^2$ in Eq.~(\ref{chi2_survey}). Marginalizing over $b_{\infty}$ leads to the 1$\sigma$ error predictions ($\Delta f_{NL}$'s) shown in Fig.~(\ref{dfnl}).

We should note that as the non-gaussian bias correction grows as $k^{-2}$ on large scales, almost all the statistical power for constraints on $f_{NL}$ comes from the largest wavelength modes in the survey, i.e. roughly $\Delta f_{NL} \propto (k_{min}/V)^{1/2}$, where $V$ is the volume of the survey. While, for our predictions, we use the rough redshift-dependent estimate:
\beq k_{min}(z) \simeq \frac{2\pi}{\left[4\pi f_{sky} r(z)^2\right]^{1/2}},\eeq
the actual constraints will depend on the large scale geometry of the survey. We also assume $k_{max}$ to be the rough scale above which the linear bias framework cannot be used reliably: \beq \frac{k_{max}^3P(k_{max},z)}{2\pi^2} \simeq 1.\eeq This choice also becomes important for marginalization over $b_{\infty}$.

Finally, for completeness, we should point out how the results of Fig.~(\ref{dfnl}) can be translated for different values of (small-scale gaussian) galaxy bias. It is easy to see that if we define:
\beq
\tilde{f}_{NL} \equiv (1-b^{-1}_{\infty}) f_{NL}, \quad \tilde{N}_{gal} \equiv b^2_{\infty} N_{gal}, \eeq
the observational constraints on $\tilde{f}_{NL}$ will only depend on $\tilde{N}_{gal}$ and the survey geometrical properties, but will be independent of the actual value of $b_{\infty}$\footnote{Here, we have ignored the bias dependence of the elliptical collapse threshold (Fig.~\ref{deltaec}), as it could only change the error predictions by $10-20\%$.}. As an example, if the mean bias of the galaxy sample is $1.5$ rather than $2$, the predicted errors will be $50\%$ larger than those in Fig.~(\ref{dfnl}) for the same survey geometry. However, to achieve this, the survey should also have more galaxies by a factor of $(4/3)^2 = 16/9$.

To summarize this section, we have shown that clustering of galaxies on large scales could provide a very sensitive smoking gun for (local) primordial non-gaussianity. For a (futuristic) deep enough galaxy survey, and in lieu of survey/atmospheric systematics, the constraint could be as good as $\Delta f_{NL} \sim 0.1$ which is an order of magnitude better than the best predicted CMB constraints (and is consistent with recent prediction of \cite{McDonald}). We will discuss more immediate future prospects in Sec. \ref{conclude}.

\section{Non-gaussianity and the Integrated Sachs-Wolfe effect}\label{isw}

In the last section, we showed that observations of clustering of galaxies on large scales could potentially lead to very sensitive constraints on a local primordial non-gaussianity. However, in order to be competitive with upcoming CMB constraints, the selection strategy of the survey needs to be uniform to 1 part in $10^5$ on the largest angles (which are most sensitive to the non-gaussian bias). Otherwise, the correlation function will be dominated by the survey systematics, which can lead to a false detection of positive non-gaussianity.

While such survey calibration seems daunting by today's standards, an alternative method that is not prone to such systematic errors is to use cross-correlation of the galaxy survey with an already well-calibrated large scale structure survey, i.e. the CMB sky. Although most of the CMB anisotropies on large angles were originated at the last scattering surface, a small fraction of them were generated after the onset of dark energy domination through the Integrated Sachs-Wolfe (ISW) effect \cite{Sachs:1967er}. It was proposed that the ISW effect could be detected through cross-correlating CMB with the galaxy surveys \cite{Crittenden:1995ak,Peiris:2000kb,Cooray:2001ab}, and its subsequent detection in cross-correlation of WMAP maps with various galaxy surveys has been interpreted as independent evidence
for the presence of dark energy \cite{Boughn:2003yz,Fosalba:2003iy,Scranton:2003in,Nolta:2003uy,Afshordi:2003xu,Padmanabhan:2004fy,Cabre:2006qm,Giannantonio:2006du,Rassat:2006kq,Ho:2008bz,Giannantonio:2008zi}.
  However, this analysis should also be sensitive to galaxy bias on large scales, and therefore (as also pointed out in \cite{Dalal:2007cu}) could be a sensitive probe of primordial non-gaussianity through the non-gaussian bias correction of the previous section. In fact, {\it both the non-gaussian contribution to the galaxy distribution and the ISW effect follow the Newtonian potential on large scales}, and therefore should be well correlated. In this section we predict the potential of this method for constraining primordial non-gaussianity.

  In the linear regime, the perturbations in the number density of galaxies of a redshift survey projected onto the sky can be written as a line of sight integral over the total matter overdensity:
  \beq
  \delta_{gal}(\hat{\bf n}) = N^{-1}\int dr r^2 n_{gal}(r) b_{gal}(r) \delta_{m}(r\hat{\bf n};t),
  \eeq
   where $n_{gal}(r)$ is the comoving mean number density of galaxies, $b_{gal}(r)$ is the average bias for the galaxy distribution, and $N$ is the mean number of galaxies per steradian. The secondary anisotropy induced in the CMB due to the ISW effect is given by:
  \beq
  \delta_{ISW}({\hat{\bf n}}) = -2\int dr \frac{d\ln g}{dr}\Phi(r\hat{\bf n};t),
  \eeq
   where  $g(t)$ is again the growth factor for the Newtonian potential.
  The angular cross-power spectrum of the projected galaxy distribution and the ISW effect is well described by the Limber approximation
   \cite{Afshordi:2003xu}:
  \bea
  &&C(\ell)_{gal,ISW} = \\ \nonumber  &&-2N^{-1}\int dr \frac{d\ln g}{dr}  n_{gal}(r) b_{gal}(r) P_{\Phi,m} \left({\ell+1/2 \over r};t\right),
  \eea
  where $P_{\Phi,m}(k;t)$ is the 3D cross-power spectrum of the Newtonian potential with the comoving matter overdensity.

 \begin{figure}
\includegraphics[width=\columnwidth]{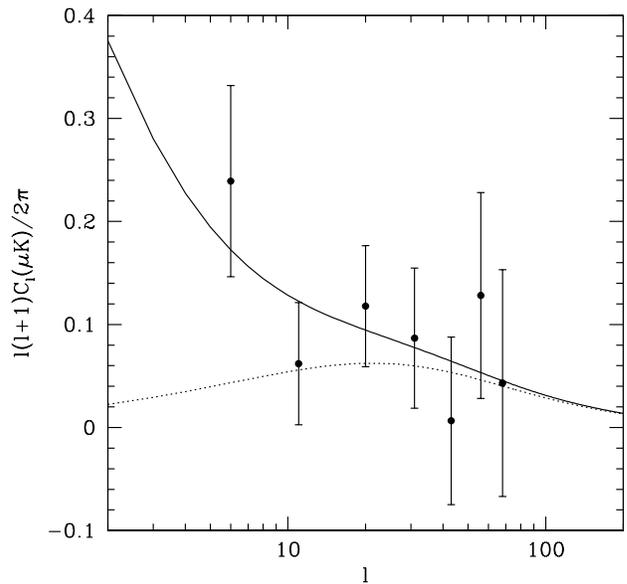}
\caption{Cross-power spectrum between NVSS radio galaxy survey and the CMB sky. The data points are from Ho et al. \cite{Ho:2008bz}. The dashed line shows the expectation from WMAP5 concordance cosmological model assuming a constant bias, while the solid line uses the best (unmarginalized) fit {\it non-gaussian} bias to the data, which corresponds to $f_{NL} = 236 \pm 127$.
  \label{nvss_isw}}
\end{figure}

 As we argued in the previous section, a local primordial non-gaussianity results in a scale dependent bias factor, $b_{gal}(k,r)$, where $k \simeq {\ell + 1/2 \over r}$ in the Limber approximation. Since the non-gaussian bias correction (Eq.~\ref{bias_ng}) grows as $k^{-2}$ on large scales, it can dominate the gaussian bias for $f_{NL} \gtrsim 1$ on scales smaller than the horizon. Therefore, the cross-correlation of the CMB and galaxy distribution can be a sensitive probe of $f_{NL}$ on large angles. As an example, Fig.~(\ref{nvss_isw}) compares the expected cross-power spectra between NVSS radio galaxy survey and the CMB sky. The data points are the recent measurements from Ho et al. \cite{Ho:2008bz}. The dashed line shows the expectation from WMAP5 concordance cosmological model and the redshift distribution provided in \cite{Ho:2008bz}, assuming a constant bias. The solid line uses the best (unmarginalized) fit {\it non-gaussian} bias to the data, which corresponds to $f_{NL} = 236 \pm 127$.

 While this result can only be cautiously interpreted as a $2\sigma$ hint for a local primordial non-gaussianity, we notice that it has the same sign (and comparable magnitude) as the current $2-3\sigma$ evidence for primordial non-gaussianity from WMAP CMB maps (see Sec. \ref{intro}), even though they correspond to different scales, eras, and measurement methods.

 Finally, we can predict how well one can measure the local non-gaussianity from cross-correlation of CMB with future galaxy surveys. To do this, we follow the method outlined in \cite{Afshordi:2004kz}, where we divide up the galaxy survey into separate spherical shells and add up the signal-to-noise for the detection of a given $f_{NL}$ in each shell in quadratures. The error for this measurement is proportional to the clustering of galaxies, and so the correlation between measurements in different shells can be safely ignored on large scales (at least for small $f_{NL}$). Within this approximation, the optimum signal-to-noise for a (full-sky) detection of a given $f_{NL}$ is given by:

  \begin{figure}
\includegraphics[width=\columnwidth]{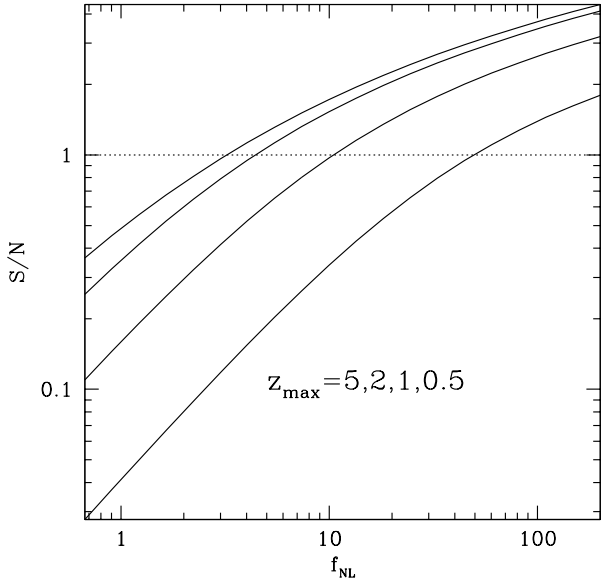}
\caption{Optimum full-sky signal-to-noise for the detection of a possible $f_{NL}$ through cross-correlation of a full-sky redshift survey (assuming $b_{gal} \simeq 2$) with the CMB sky. The four curves correspond to the maximum redshift of the survey, $z_{max} = 0.5,1,2,$ and $5$.   \label{chi2}}
\end{figure}
 \bea
 &&(S/N)^2 \simeq  \sum_{\ell} \frac{4(2\ell+1)}{C_{TT}(\ell)}\times \\ \nonumber &&\int \frac{dr(z)}{r(z)^2} \left(\frac{d\ln g}{dr}\right)^2 \left(\Delta b_{gal}(k,z)\over b_{gal}(k,z)\right)^2 \frac{P^2_{\Phi,m}(k,z)}{P(k,z)}, \label{s_n}\eea
where $C_{TT}(\ell)$ is the angular auto-power spectrum of the CMB, and $P(k,z)$ is the linear matter power spectrum. Moreover, $b_{gal}$ and $\Delta b_{gal}$ ($\propto f_{NL}$) are the total bias and its non-Gaussian correction (obtained in Eq.~\ref{bias_ng}) respectively. As in the Limber approximation, $k \simeq {\ell + 1/2 \over r}$ in Eq.~(\ref{s_n}).

Fig.~(\ref{chi2}) shows the predicted signal-to-noise (from Eq.~(\ref{s_n})) for the detection
of primordial non-gaussianity through cross-correlation with the CMB for a full-sky redshift survey that extends out to $z_{max} = 0.5,1,2,$ and $5$, where we assumed the small-scale gaussian bias $b_{gal} \simeq 2$ for concreteness. Therefore, we predict that cross-correlation of CMB and large scale structure surveys can eventually be competitive with the CMB surveys in providing independent constraints on local primordial non-gaussianity at the level of $\Delta f_{NL} \sim 3$.

In this calculation, we have ignored the Poisson noise (due to a finite number of galaxies) or other possible survey systematics. However, we should note that since these errors do not contribute to the ISW signal \footnote{A notable exception is Galactic foregrounds in the CMB which could be correlated with Galactic contaminations of galaxy surveys. However, similar to the analysis of primary CMB anisotropies \cite{Gold:2008kp}, the frequency dependence of these foregrounds can provide a way to clean the cross-correlation signal from Galactic contaminations.}, they cannot cause a systematic bias in $f_{NL}$ determinations from ISW-galaxy cross-correlation analyses. Nevertheless, Poisson noise/survey systematics can increase the random error in cross-correlation, which is manifested through increasing the apparent auto-correlation of the galaxy survey. Since we have already calculated the sensitivity of the galaxy auto-correlation function to $f_{NL}$ in the last section (Fig.~(\ref{dfnl})), we can say that assuming $f_{NL} \gtrsim 3$, for $N_{gal} \gtrsim 10^{7-8}$ the galaxy auto-correlation will become roughly insensitive to the Poisson noise on large scales (where the non-gaussian signal dominates). Therefore, the determination of $f_{NL}$ from ISW-galaxy cross-correlation will not significantly suffer from Poisson noise, as long as the number of galaxies in a (full-sky) survey is larger than $10^{7-8}$. Assuming that survey systematics also look similar to Poisson noise on large angles, this also implies that the calibration of the galaxy survey selection function must be better than
$1/\sqrt{10^8} = 10^{-4}$, in order not to affect the sensitivity of non-gaussianity measurements.

%\pacs{13.15.+g, 64., 64.30.+t, 64.70., 64.70.Fx, 98.80., 98.80.Cq}

\section{Conclusions and Future Prospects}\label{conclude}

In this paper, we have studied the impact of of a local primordial non-gaussianity (expected from super-horizon evolution in multi-field inflationary models or the new ekpyrotic scenario) on the statistics of collapsed objects in the late universe. We developed an approximation that describes the linear non-gaussian density field as a local bilinear combination of gaussian fields. Applying this to the elliptical collapse framework enabled us to find an analogue gaussian linear density field, which should produce the same predictions as the non-gaussian system for collapsed objects of a given mass, assuming small primordial non-gaussianity. For halo mass function, this prescription leads to an effective analytic correction to $\sigma(M)$, the amplitude of linear fluctuations on mass scale $M$, which can be readily applied to Press-Schechter-like analytic approximations for the mass function. We found this to give roughly consistent results with direct simulations with non-gaussian initial conditions. We then focused on the impact of non-gaussianity on the clustering of galaxies, and confirmed the recent discovery of a non-local bias correction associated with local non-gaussian initial conditions \cite{Dalal:2007cu,Matarrese:2008nc}. We found that auto-correlation of galaxy surveys can ultimately lead to a precision $\Delta f_{NL} \sim 0.1$ which is an order of magnitude better than the best anticipated CMB constraints. Cross-correlation of CMB and galaxy surveys (through the ISW effect), even though suffering from a low signal-to-noise due to primary CMB contamination, is less prone to survey systematics, and can also yields $\Delta f_{NL} \sim 3$, which is again competitive with future CMB constraints. Interestingly, the cross-correlation of CMB and the NVSS radio galaxy survey already provides a $2\sigma$ hint for a large primordial non-gaussianity $f_{NL} = 236 \pm 127$.

Finally, we should comment on the future developments of the studies of primordial non-gaussianity. While five more years of WMAP observations can only marginally improve the current constraints to $\Delta f_{NL} \sim 15-20$, the most anticipated observational improvement is the first data release of Planck satellite, expected in 2012. After careful accounting for CMB secondary anisotropies, as well as the 2nd order corrections to the CMB Boltzmann solvers, Planck is expected to achieve $\Delta f_{NL} \sim 3$. However, as we demonstrated in this paper, competitive or even superior constraints can come from the study of clustering of large scale galaxy surveys. For example, Large Synoptic Survey Telescope (LSST) is expected to survey half of the sky down
to magnitude $r\sim 27.5$ finding $\sim 10^{10}$ galaxies out to $z_{max} \sim 2$. Using Fig.~(\ref{dfnl}), LSST can yield a precision of $\Delta f_{NL} \sim 2-3$ through its galaxy auto-correlation function. The cross-correlation of LSST with the CMB sky could independently yield $\Delta f_{NL} \sim 7$. Similar proposed or underway deep, large area surveys such as Pan-STARRS, ADEPT, CIP, etc. could potentially lead to comparable constraints. However, one important point to note is that {\it precise spectroscopic redshifts are not required for non-gaussianity determination, as most of the non-gaussian signal comes from the largest scales of the survey}.

Shortly before the submission of this paper, Slosar et al. \cite{Slosar:2008hx} published a comprehensive study of this effect in the present large-scale cosmological galaxy surveys, leading to a significant constraint: $f_{NL} = 28 \pm 25$. Nearly all the statistical weight of this constraint is based on the lack of anomalous power in the lowest l-bin of the auto-power spectrum of a high redshift sample of $\sim 10^5$ SDSS photometric QSO's, which cover $15\%$ of sky and have an effective mean bias of $\sim 2.75$. Plugging these numbers into the error estimates of Sec. \ref{clustering} (Fig. \ref{dfnl}), we would predict an error of about a factor of $2$ worse than what was found in \cite{Slosar:2008hx}. As we pointed out in Sec. \ref{clustering}, the actual error on $f_{NL}$ will be sensitive to the survey geometry, and so this disagreement between our rough error prediction and Slosar et al.'s findings may not be an immediate cause for alarm. Nevertheless, it is curious that despite the lack of any redshift information (which was used in our error prediction), as well as possible contaminations/survey systematics, Slosar et al. could achieve an accuracy in $f_{NL}$ comparable or better than an optimistic error prediction. A further concern is that the correlated Poisson noise (typically expected from non-linear clustering, e.g. \cite{McDonald:2006mx}) is not marginalized over in their analysis.

On a different note, Slosar et al. point out that a degeneracy between $f_{NL}$ and the redshift and bias distribution of NVSS survey (which are both poorly known), hinders any accurate determination of $f_{NL}$ from NVSS data, or its correlation with the CMB. While, at face value, this might seem inconsistent with our $\sim 2\sigma$ evidence for non-gaussianity from the correlation of NVSS with WMAP, we should note that both redshift and bias distributions may eventually be fixed, either from theoretical modeling, or direct observations, which can then realize the full statistical power of cross-correlation constraints.

On the theoretical front, we are still far from a good theoretical understanding of non-linear structures from non-gaussian initial conditions. While the present work provides a bridge between this problem, and the huge body of work on non-linear/collapsed structures from gaussian initial conditions (such as the semi-analytic frameworks, the halo model, as well the N-body/hydro numerical simulations), it does suffer from some simplifying assumptions. The first is the ellipsoidal collapse scenario \cite{Sheth:1999su, Sheth:2001dp}. Although for large masses and $f_{NL}\epsilon(M) \ll 1$ (i.e. near-gaussian initial conditions), the collapse should be well-described as spherical (leading to a universal spherical collapse threshold), for smaller masses, the threshold depends on the ellipticity of the proto-halo, which is a random number with a distribution set by the statistics of the initial conditions. The current analytic form (Eq.~\ref{ecol}) is a fit to random realizations of gaussian fields with $\Lambda$CDM power spectrum, and thus there is no reason to exclude corrections linear in $f_{NL}$, if the linear density field obeys non-gaussian statistics. Therefore, one clear direction for improvement is to investigate/extend the ellipsoidal collapse framework for non-gaussian initial conditions.

The second simplification (and probably the most non-trivial part of this work) is the local approximation of Eq.~(\ref{deltax}), which was instrumental in all the subsequent analysis/predictions. While the consistency of our results for mass function and clustering with other numerical and analytic studies \cite{Dalal:2007cu,Matarrese:2008nc} verifies the accuracy of this approximation, Eq.~(\ref{deltax}) could also be viewed as a {\it convenient parametrization for the primordial non-gaussianity}. To see this, note that for an $f_{NL}$-type non-gaussianity (Eq.~\ref{fnl}), the bispectrum of $\Phi$, $B(k_1,k_2,k_3)$,  is written in terms of the power spectrum of the Gaussian field $\Phi_{pG}$, $P_{pG}(k)$:
\bea
B(k_1,k_2,k_3) = -2f_{NL}P_{pG}(k_1)P_{pG}(k_2)P_{pG}(k_3)\times\nonumber\\ \left[P^{-1}_{pG}(k_1)+P^{-1}_{pG}(k_2)+P^{-1}_{pG}(k_3)\right] \label{bispectrum},
\eea
where $P_{pG}(k) \propto k^{-3}$ for scale-invariant spectra.
While our local approximation does not exactly preserve this form, different observables are mostly sensitive to a limited region in the $k_1,k_2,k_3$ space, where Eq.~(\ref{deltax}) may provide a simple parametrization. For example, it is reasonable to think that measurements of mass function of haloes of mass $M$ would only be sensitive to non-gaussianity around $k_1 \sim k_2 \sim k_3 \sim \pi/R(M)$ (see the line preceding Eq.~\ref{epsilon}). Therefore, constraints on $f_{NL}$ using mass function could be roughly translated into constraints on the bispectrum for {\it near-equilateral triangles} of the side $\sim \pi/R(M)$.

A more interesting case is the clustering (or the power spectrum) of collapsed haloes (e.g. galaxies) that we studied in Secs. \ref{clustering}-\ref{isw}. In this case, we are dealing with a hierarchy of scales: $k_3 \ll k_1,k_2$, where $k_1 \ll 0.01~ \mpc^{-1}$ is the scale over which we measure the galaxy power spectrum, and as we showed pre-dominantly corresponds to the modes of the size of the survey. On the other hand, $k_1,k_2 \sim \pi/R(M) \gtrsim 0.1 ~\mpc^{-1}$ correspond to the comoving size of the collapsing regions. Therefore, observations of the
galaxy power spectrum provides an exquisite measure of the bispectrum for {\it squeezed triangles}. In particular, the non-gaussian bias correction $\Delta b_{NG}(k_3) \propto k_3^{-2}$ found in Sec. \ref{clustering} (as well as in  \cite{Dalal:2007cu,Matarrese:2008nc}) are specific to the local form of non-gaussianity (Eq.~\ref{bispectrum}) which is only produced through super-horizon evolution, where $B(k_1,k_2,k_3) \propto k_3^{-3}$ as $k_3 \rightarrow 0$. In contrast, for non-gaussianities produced through sub-horizon evolution (such as in slow-roll single field inflation, DBI, ghost inflation and k-inflation), $B(k_1,k_2,k_3) \rightarrow $ const. or 0 in this limit \cite{Babich:2004gb}. Therefore, by way of comparison,  the scale-dependent bias $\Delta b_{NG}(k_3) \rightarrow 0$ (and so is unobservable) in the long wavelength limit, for models that produce equilateral primordial non-gaussianity.

We thus close this paper by the exciting (although, by now, familiar) premise of the cosmological microscope as a probe of the fundamental physics of the early universe. In particular, that an anomalously large clustering of galaxies on large scales will be a unique smoking gun for non-linear super-horizon evolution (and thus large primordial isocurvature modes) during the early stages of cosmogenesis.
% gaussian statistics for power spectrum?

 We would like to thank Neal Dalal for providing us with the mass functions from their non-gaussian simulations \cite{Dalal:2007cu}. We also would like to thank Neal Dalal, Chris Hirata, Shirley Ho, Marilena LoVerde, Pat McDonald, Bob Scherrer, Uros Seljak, Anze Slosar, David Spergel, Michael Strauss, and James Taylor for useful comments and illuminating discussions.  This research was supported by Perimeter Institute for Theoretical Physics.  Research at Perimeter Institute is supported by the Government of Canada through Industry Canada and by the Province of Ontario through the Ministry of Research \& Innovation.

\bibliographystyle{utcaps_na2}
\bibliography{gauss_cluster}

\end{document}